\documentstyle[aaspp4]{article}

\font\sevenrm=cmr7
\def\cosmo{H$_0$ = 50~km~s$^{-1}$~kpc$^{-1}$ and q$_0 = 0$} 
\def\en{\enspace}

\input{psfig.tex}

\received{July 14, 1998}
\accepted{August 20, 1998}
\journalid{Astrophysical Journal}{}

\begin{document}

\slugcomment{\centerline{Accepted on August 20, 1998 for publication 
by the {\it Astrophysical Journal}}}

\title{\bf HST Observations of the Host Galaxies of BL~Lacertae Objects}

\author{C. Megan Urry} 
\authoraddr{Space Telescope Science Institute, 3700 San Martin Dr., Baltimore,
MD 21218, USA}
\authoremail{cmu@stsci.edu}
\affil{Space Telescope Science Institute}

\author{Renato Falomo} 
\authoraddr{Osservatorio Astronomico di Padova, Vicolo dell'Osservatorio 5, 
35122 Padova, Italy}
\authoremail{falomo@astrpd.pd.astro.it}
\affil{Astronomical Observatory of Padova}

\author{Riccardo Scarpa\altaffilmark{1}}
\authoraddr{Space Telescope Science Institute, 3700 San Martin Dr., 
Baltimore, MD 21218, USA}
\authoremail{scarpa@stsci.edu}
\affil{Space Telescope Science Institute}
\altaffiltext{1}{also at Department of Astronomy, Padova University, Vicolo 
dell'Osservatorio 5, 35122 Padova, Italy}

\author{Joseph E. Pesce} 
\authoraddr{Department of Astronomy, Pennsylvania State University,
525 Davey Lab, University Park, PA 16802 }
\affil{Pennsylvania State University}
\authoremail{pesce@astro.psu.edu}

\author{Aldo Treves} 
\authoraddr{University of Milan at Como, Via Lucini 3, Como Italy}
\authoremail{treves@uni.mi.astro.it}
\affil{University of Milan at Como}

\author{Mauro Giavalisco\altaffilmark{2}}
\authoraddr{Carnegie Observatories, 813 Santa Barbara St., 
Pasadena CA 91101-1292}
\authoremail{mauro@ociw.edu}
\affil{Carnegie Observatories}
\altaffiltext{2}{Hubble Fellow}

\begin{abstract}

Six BL~Lac objects from the complete 1~Jy radio-selected sample of
34 objects
were observed in Cycle~5 with the {\it HST} WFPC2 camera to an equivalent
limiting flux of $\mu_I \sim26$~mag~arcsec$^{-2}$.
Here we report results for the second half of this sample, as well as
new results for the first three objects, discussed previously by 
Falomo et al. (1997).
In addition, we have analyzed
in the same way {\it HST} images of three X-ray-selected BL~Lacs observed 
by Jannuzi et al. (1997).
The ensemble of 9 BL~Lac objects
spans the redshift range from $z=0.19$ to $\sim1$.
Host galaxies are clearly detected in seven cases, while the
other two, at $z\sim0.258$ (redshift highly uncertain) and $z=0.997$,
are not resolved. 
The {\it HST} images constitute a homogeneous data set with
unprecedented morphological information between a few tenths of 
an arcsecond and several arcseconds from the nucleus, 
allowing us in 6 of the 7 detected host galaxies to rule out 
definitively a pure disk light profile.
The host galaxies are luminous ellipticals with an average absolute
magnitude of $M_I \sim -24.6$~mag (with dispersion 0.7~mag), 
more than a magnitude brighter
than $L^*$ and comparable to brightest cluster galaxies.
The morphologies are generally smooth and have small 
ellipticities ($\epsilon\lesssim 0.2$). 
Given such roundness, there is no obvious 
alignment with the more linear radio structures. In the six cases 
for which we have {\it HST} WFPC2 images in two filters, 
the derived color profiles show no strong spatial gradients and are as
expected for K-corrected passively evolving elliptical galaxies.
The host galaxies of the radio-selected and X-ray-selected BL~Lacs
for this very limited sample are comparable in both morphology and 
luminosity.

{\underline{\em Subject Headings:}}
BL~Lacertae objects: individual (0814+425, 0828+493, 1221+245, 1308+326,
1407+595, 1538+149, 1823+568, 2143+070, 2254+074) ---
galaxies: structure ---
galaxies: elliptical

\end{abstract}

\section{Introduction}
\label{sec:intro}

AGN are known to lie in galaxies and indeed both galaxies and AGN have
evolved similarly, in terms of number and luminosity, over cosmic time
(e.g., Silk \& Rees 1998). This suggests a close connection between
phenomena associated with black holes and the formation and evolution
of ordinary galaxies. Thus it is of considerable interest to understand
the properties of AGN host galaxies themselves --- their morphology/type, 
luminosity, color --- and to explore their
similarities and differences with respect to normal galaxies. 

Studies of AGN host galaxies are hampered by the technical difficulty of
detecting host galaxy light beneath the bright, seeing-smeared AGN light.
The order of magnitude improvement in spatial resolution possible with 
the {\it Hubble Space Telescope (HST)} has made an important difference,
providing unique and critical information at subarcsecond scales.
{\it HST} images provide more detailed information on nearby galaxies,
such as the ability to distinguish between elliptical and disk light 
distributions, and allow more distant host galaxies to be resolved for
the first time.

Other {\it HST} programs have focused largely on quasars, many at $z<0.4$ 
(Bahcall et al. 1997;
Disney et al. 1995;
McLeod \& Rieke 1995;
Hooper, Impey \& Foltz 1997)
and also a number at higher redshift, $z\gtrsim 1.5$,
with more recent observations
filling in the range from $z=0.4$ to $\sim2$
(Best, Longair \& R\"ottgering 1997;
Ridgeway \& Stockton 1997;
Serjeant, Rawlings \& Lacy 1997).
These include both 
radio-loud and radio-quiet quasars. A number of radio galaxies 
have also been observed with {\it HST} (McCarthy et al. 1997). 
The origin of the radio-loudness distinction is still unclear --- 
it could be due to internal properties or to environmental influences 
or even to selection effects (Blandford 1990, Wilson \& Colbert 1995, 
Smith et al. 1986, Perlman et al. 1998).
We have chosen to concentrate on
radio-loud AGN, which seem less affected by dense gaseous environments.
These are increasingly well understood as a unified family of 
black-hole-powered AGN with powerful relativistic jets, with subclasses
related through orientation and luminosity (Urry \& Padovani 1995;
Sambruna, Maraschi \& Urry 1996; Fossati et al. 1997).

Our {\it HST} program\footnote{Based on observations made with the NASA/ESA
Hubble Space Telescope, obtained at the Space Telescope Science
Institute, which is operated by the Association of Universities for
Research in Astronomy, Inc., under NASA contract NAS~5-26555.}
targeted radio-loud AGN with intrinsically 
lower power radio jets, typical of the Fanaroff-Riley type I radio galaxies,
so as to explore the interplay of AGN luminosity and host galaxy
properties. 
According to unified schemes, the aligned version of low-luminosity
(FR~I-type) radio galaxies are BL~Lac objects, while
quasars are the aligned version of high-luminosity
(FR~II-type) radio galaxies (Urry \& Padovani 1995). 
Because of their orientation, BL~Lac radiation is beamed toward us,
so they can be found easily to much higher redshifts than even quite luminous
radio galaxies. Thus our {\it HST} observations of
BL~Lac objects extend host galaxy studies significantly
in the $L-z$ plane.

Results for three objects from our
Cycle~5 sample of radio-selected BL~Lacs
were reported previously by
Falomo et al. 1997 (hereafter Paper~I).
Here we report the results for the remaining three objects and give 
some new results for the first three.
We also present a uniform analysis of the {\it HST} images of
the three X-ray-selected BL~Lacs discussed by
Jannuzi et al. (1997).

The present sample of 9 objects represents the best data 
available for BL~Lacs with redshifts $z\gtrsim0.5$ 
(i.e., 5 of the present sample of 9 objects, 
including 4 LBL and 1 HBL).\footnote{As is the case for many BL~Lac objects,
the redshifts of 5 of our 9 targets are somewhat uncertain.
For 0814+425, the reported redshift $z=0.258$ is almost certainly
not correct.
It is based on weak detection of
Mg~{\sevenrm II}~$\lambda 2798$ and [O~{\sevenrm II}]~$\lambda 3727$
emission lines (Wills \& Wills 1976) which were not seen in a second
spectrum by the same authors or in independent 
published spectra (Dunlop et al. 1989; Stickel, Fried \& K\"uhr 1993).
Most recently Lawrence et al. (1996) obtained a very high signal-to-noise 
spectrum, which shows three weak features at either $z=0.245$ or 1.25, but
in any case not at $z=0.258$. The other redshifts
are more likely correct but still not firm. 
The values for 1538+149 ($z=0.605$; Stickel et al. 1993) and
1823+568 ($z=0.664$; Lawrence et al. 1986) are based on
identification of two weak emission lines as
Mg~{\sevenrm II} and [O~{\sevenrm II}], and
for 0828+493 ($z=0.548$) a single emission line
was identified as O~{\sevenrm II}~$\lambda3727$ (Stickel et al. 1993).
Finally, the redshift of 1221+245 ($z=0.218$; Stocke et al. 1991)
depends on Ca~{\sevenrm II} and Mgb features that are of marginal
significance and do not line up perfectly (Morris et al. 1991).} 
We concentrate here on the issue of host galaxies as observed with 
{\it HST}. The environments of all 9 BL~Lac objects, including the
presence (and in some cases, colors) of close companions, will be
reported in a separate paper (Pesce et al., in preparation). 
The observations
and data analysis are described in \S~\ref{sec:obsda}
and the results for individual
objects in \S~\ref{sec:results}, 
along with comparisons to previous measurements of
the host galaxy where appropriate. The implications 
are discussed and conclusions summarized in \S~\ref{sec:disc}.
Throughout the paper we used \cosmo. 

\section{Observations and Data Analysis}
\label{sec:obsda}

\subsection{The Observed BL~Lac Samples}
\label{ssec:sample}

Current samples of BL~Lacertae can be divided empirically into two
types, ``red'' or ``blue'', depending on their spectral energy distributions.
The ``red'' or low-frequency-peaked BL~Lacs (LBL)
have peak synchrotron powers at infrared-optical wavelengths
and have been found in radio-selected surveys.
The ``blue'' or high-frequency-peaked BL~Lacs (HBL) 
peak at UV-X-ray wavelengths and dominate current X-ray-selected
samples.
(There is probably a continuous distribution
of synchrotron peaks between the IR and X-ray but current selection 
techniques emphasize the extreme ends of the distribution;
Laurent-Muehleisen 1997.)
There are significant trends in spectral
shape with luminosity, with the ``red'' objects being systematically
higher in luminosity
than the ``blue'' (Sambruna et al. 1996, Fossati et al. 1997). Clearly
the range of spectra results from range in jet physics (Ulrich et al. 1997)
and therefore represents an interesting diversity of AGN energetics. 

Since the ``blue'' objects in current samples have lower redshifts
and so can be relatively well-studied from the ground,
our Cycle~5 {\it HST} project focused on the higher redshift 
``red'' objects. The targets were taken from the complete
1~Jy sample of 34 BL~Lac objects by Stickel et al. (1991) after restricting the
redshift range to $z\gtrsim 0.2$, since nearer objects are easily resolved from
the ground, and to $z\lesssim1$, to ensure detections or useful limits 
for the host galaxies in 2-orbit exposures.

A similar program in the same Cycle (PI, Buell Jannuzi) observed 
three ``blue'' BL~Lac objects (HBL), two at $z\sim 0.2$ and one $z\sim 0.5$
(Jannuzi et al. 1997, hereafter J97). 
We included these in our analysis in order to
look for systematic differences in the host galaxy properties. The
comparison to the results from J97 also offers a 
useful estimate of the systematic uncertainties in determining galaxy size
and luminosity.

\subsection{Observations}
\label{ssec:obs}

Observations were carried out as described in Paper~I and are only
briefly reviewed here.
All objects were observed with the {\it HST} WFPC2 camera in
the F814W filter. To obtain for each target a final image well exposed
both in the inner, bright nucleus and in the outer regions where the
host galaxy emission dominates, we set up a
series of several pairs of exposures with duration ranging from a few tens to
500-1000 seconds. 

Each BL~Lac object was observed for a total of two
orbits, with each orbit consisting of the same run of exposure
times. Between orbits, we rolled the telescope by $\sim20^\circ$, so
that the diffraction spikes rotated relative to the sky. This
technique allows us to detect easily any asymmetric emission (e.g.,
jets, distortions, companions) independent of contamination by the
diffracted nuclear light. 
The dates of the {\it HST}
observations, total exposure times, number of exposures, and roll
angles are given in Table~1. 
Also reported in this Table are the 
{\it HST} Cycle~6 snapshot observations used to derive color profiles
(Urry et al., in preparation).
Finally, we include in Table~1 the estimated sky magnitude and rms uncertainty
for each Cycle~5 observation, computed from source-free regions over the
full PC image. The quantity $\sigma_{sky}$ therefore includes both 
statistical error and the contribution 
from large-scale fluctuations in the ``sky'' background (e.g., variations 
introduced by imperfect flat fielding).

\begin{deluxetable}{ccccccccc}
\footnotesize
\tablewidth{5.7in}
\tablecaption{Journal of HST Observations}
\tablehead{
\colhead{Object} & \colhead{z}
        & \colhead{$N_H$\tablenotemark{(a)}}
        & \colhead{Date} & \colhead{Filter\tablenotemark{(b)}}
        & \colhead{Exptime}
        & \colhead{No.Exp.}
        & \colhead{Roll\tablenotemark{(c)}}
        & \colhead{$\mu_{sky}$\tablenotemark{(d)}} \\
\colhead{} & \colhead{}
        & \colhead{($\times10^{20}$cm$^{-2}$)}
        & \colhead{} & \colhead{} & \colhead{(sec)} & \colhead{}
        & \colhead{($^\circ$)}
        & \colhead{(mag/arcsec$^2$)}
}
\startdata
0814+425 & (0.25?) & 4.92 & 13 Nov 95  &F814W& 1960 &\en6 &\en90
        &$21.8 \pm 0.2$ \nl
        & & &                          &F814W& 1960 &\en6 & 110 & \nl
0828+493 & 0.548   & 3.94 & 11 Feb 96  &F814W& 2060 &\en6 & 330
        &$21.8\pm0.2$   \nl
        &          &      &            &F814W& 2060 &\en6 & 310 & \nl
        &          &      & 27 Feb 96  &F702W&\en840&\en3 & 316 & \nl
1221+245 & 0.218   & 1.74 & 13 Jan 96  &F814W& 1900 &\en5 &\en92
        &$21.84\pm0.05$ \nl
1308+326 & 0.997   & 1.08 & 04 Mar 96  &F814W& 1800 &\en6 &\en30
        &$22.0\pm0.4$   \nl
        &          &      &            &F814W& 1800 &\en6 &\en50 & \nl
1407+595 & 0.495   & 1.55 & 22 Oct 95  &F814W& 4080 &\en7 & 182
        &$21.88\pm0.05$ \nl
        & &               & 23 May 97  &F606W&\en840&\en3 & 334  & \nl
1538+149 & 0.605   & 3.23 & 21 Mar 96  &F814W& 1670 &\en6 & 100
        &$22.0\pm0.25$  \nl
        &          &      &            &F814W& 1670 &\en6 &\en80 & \nl
        &          &      & 21 May 96  &F606W&\en440&\en4 & 355  & \nl
1823+568 & 0.664 & 4.20 & 02 Oct 95  &F814W& 2120 &\en6 & 275
        &$22.3\pm0.3 $  \nl
        &          &      &            &F814W& 2120 &\en6 & 255  & \nl
        &          &      & 30 Dec 96  &F606W&\en840&\en3 & 176  & \nl
2143+070 & 0.237   & 4.87 & 08 Sep 95  &F814W&\en850&\en7 & 303
        &$21.77\pm0.13$ \nl
        &          &      & 13 May 97  &F606W&\en500&\en3 &\en76 & \nl
2254+074 & 0.19\en & 5.39 & 12 Oct 95  &F814W& 1520 & 10  & 260
        &$21.86\pm0.04$ \nl
        &          &      &            &F814W& 1520 & 10  & 280  & \nl
        &          &      & 13 May 97  &F606W&\en320&\en4 &\en72 & \nl
\enddata
\tablenotetext{(a)}{Galactic HI column density (Stark et al. 1992),
which can be translated to interstellar extinction via
the relation derived for halo stars,
$\log{E(B-V)} = \log N_{HI} - 21.83 ~ {\rm cm}^{-2}{\rm mag}^{-1}$
(Shull \& Van Steenberg 1985).}
\tablenotetext{(b)}{Pointed observations in F814W filter;
snapshot observations in F702W or F606W.}
\tablenotetext{(c)}{Telescope roll angle in degrees.}
\tablenotetext{(d)}{I-band surface brightness (in mag/arcsec$^2$)
of sky and rms uncertainty. Value quoted refers to the average of
images at two roll angles. Sky values for F702W and F606W snapshot
images are given by Urry et al. (1998).}
\end{deluxetable}
\normalsize

\subsection{Data Reduction}
\label{ssec:datared}

After preliminary reduction (flat--field, dark and bias
subtraction, and flux calibration) carried out as part of the standard
{\it HST} pipeline processing, we combined images at the two telescope
roll angles as described in Paper~I. The final summed images of the 
three new ``red'' BL~Lac objects 
are shown in the upper panel of Figure~1.

\begin{figure}
\psfig{figure=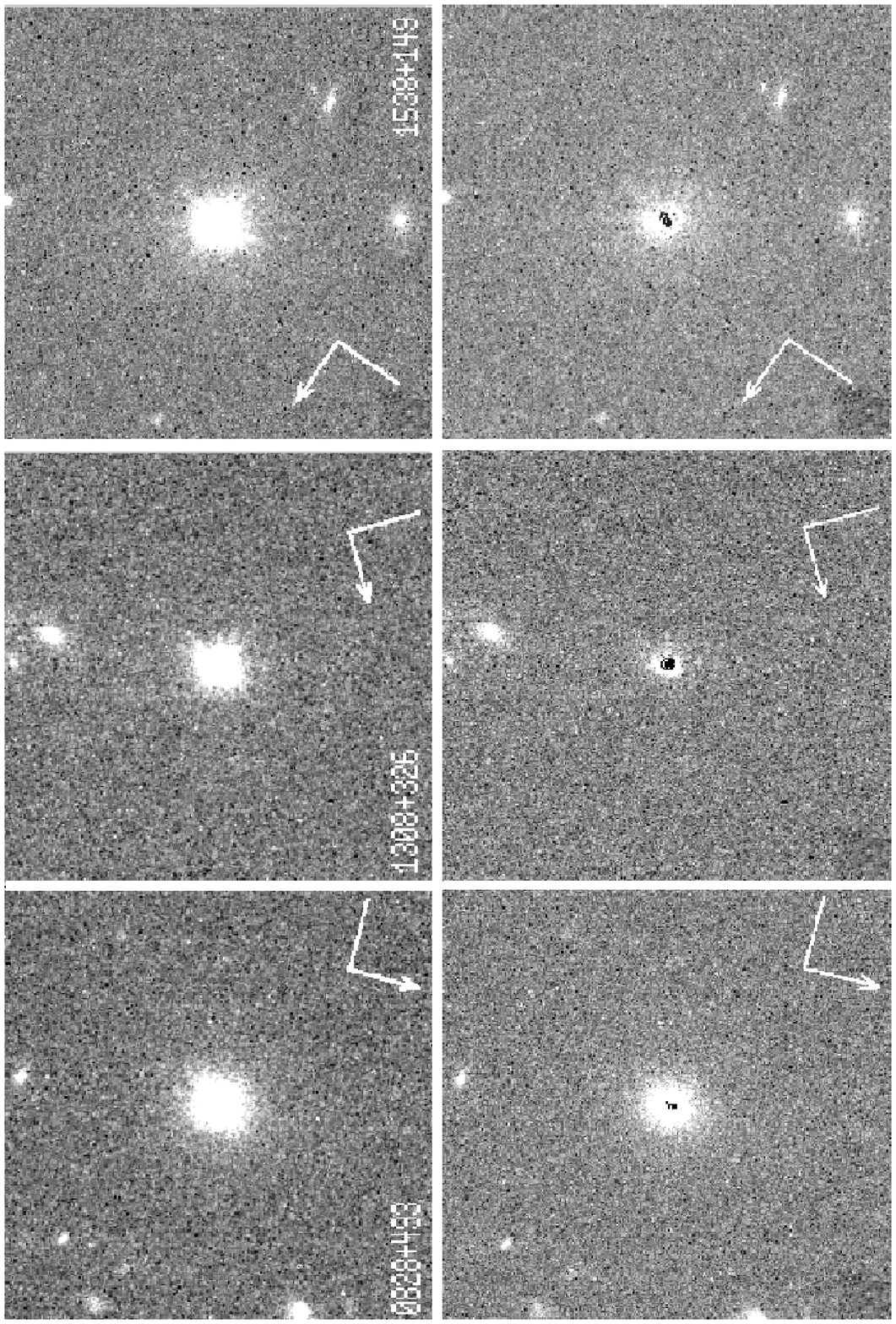,height=7.15in}
\caption{\footnotesize{\it HST} WFPC2 images of three radio-selected 
BL~Lac objects (central 300$\times$300 pixels,
or 13\farcs80$\times$13\farcs80, of the PC). 
The top panel is the summed data
(including point source). In the lower panel, the best-fit PSF has
been subtracted; residual structure in the center is due to
imperfect knowledge of the PSF.
Host galaxies are detected in 
0828+493 ($z = 0.548$) and 1538+149 ($z = 0.605$) 
but not in 1308+326 ($z = 0.997$). 
North is indicated by the arrow and East by the bar;
the length of the arrow is 2.67~arcsec, which corresponds to 
a physical length of 22.6~kpc for 0828+493, 29.0~kpc for 1308+326, 
and 21.4~kpc for 1538+149.
\normalsize}
\end{figure}

The I~magnitudes were calibrated
using the prescription of Holtzman et al. (1995, their Eq.~8 and Table~7) 
for the ``flight system'':
\begin{equation}
I = -2.5 \log ({\rm DN} ~ {\rm s}^{-1}) + K_I ,
\end{equation}
\noindent
where $K_I$ is:
\begin{equation}
K_I = 20.839 - 0.112 (R - I) + 0.084 (R - I)^2 
	+ 2.5 \log({\rm GN}) + 0.1 ~ {\rm mag} .
\end{equation}
\noindent
The constant 0.1~mag term corrects for the
infinite aperture, and GN=2 for gain=7 (our case).
The $R-I$ color assumed was for a redshifted elliptical galaxy,
interpolated from the values in Fukugita et al. (1995). 
For a typical elliptical galaxy with rest-frame color 
$R-I=0.8$, $K_I=21.65$.
This calibration is slightly different from that used in Paper~I,
where no aperture correction (0.1 mag) was applied and a fixed color 
($R - I = 0.8$) was assumed. Because the Cousins I and {\it HST} F814W 
filters are very similar, the dependence on color is always small;
for the host galaxies detected here ($z\lesssim 0.7$), 
the color terms are smaller than $\pm 0.04$~mag.
To have a complete homogeneous set of results, we report in this paper
slightly revised numbers for the three objects discussed in Paper~I.

In neither paper do we make reddening corrections for interstellar 
absorption; in all cases, the column densities are low 
($N_H \lesssim 5\times10^{20}$~cm$^{-2}$; see Table~1) 
so the effect is small, with $E(B-V) \lesssim 0.1$ using the conversion
$\log N_H/E(B-V)=21.83$~cm$^{-2}$~mag$^{-1}$ appropriate at 
high latitudes (Shull \& Van Steenberg 1995). For a total-to-selective
extinction of $A_V = 3.1 E(B-V)$, V-magnitudes for 0814+425,
2143+070, and 2254+074 could be as much as 0.3~mag larger 
($\sim0.1$~mag or less for the rest);
the much smaller effect in the I-band is well within our estimated
systematic uncertainty of a few tenths of a magnitude
(\S~\ref{ssec:systematic}).

\subsection{The Point Spread Function Used}
\label{ssec:psf}

To determine if an AGN host galaxy is resolved requires knowing 
the shape of the {\it HST} point spread function (PSF). 
In general, the PSF is not known precisely because it
is a function of position on the CCD, filter used, 
epoch of observation, telescope focus (including breathing), and so on.
The Tiny Tim model (Krist 1995) gives an excellent representation of the PSF
shape within $\sim 2$~arcsec from the center. Outside this range, 
particularly in the PC chip,
there is a substantial contribution due to scattered light, 
which is not included in the Tiny Tim model.

Since the scattered light has not been modeled analytically we derived an 
empirical profile from PC images of 3-5 very bright stars in each of
three different filters (F814W, F702W, and F606W). 
The amount of scattered light is roughly constant from image to image 
and largely insensitive to the star position in the CCD or 
to the filter used (as found by Krist \& Burrows 1994). 
Figure~2 shows the Tiny Tim and stellar PSFs for the F702W filter, 
for which the largest number of useful stars were available. 
The stellar PSF lies systematically above the Tiny Tim model 
beyond $\sim 2$~arcsec.

\begin{figure}
\psfig{figure=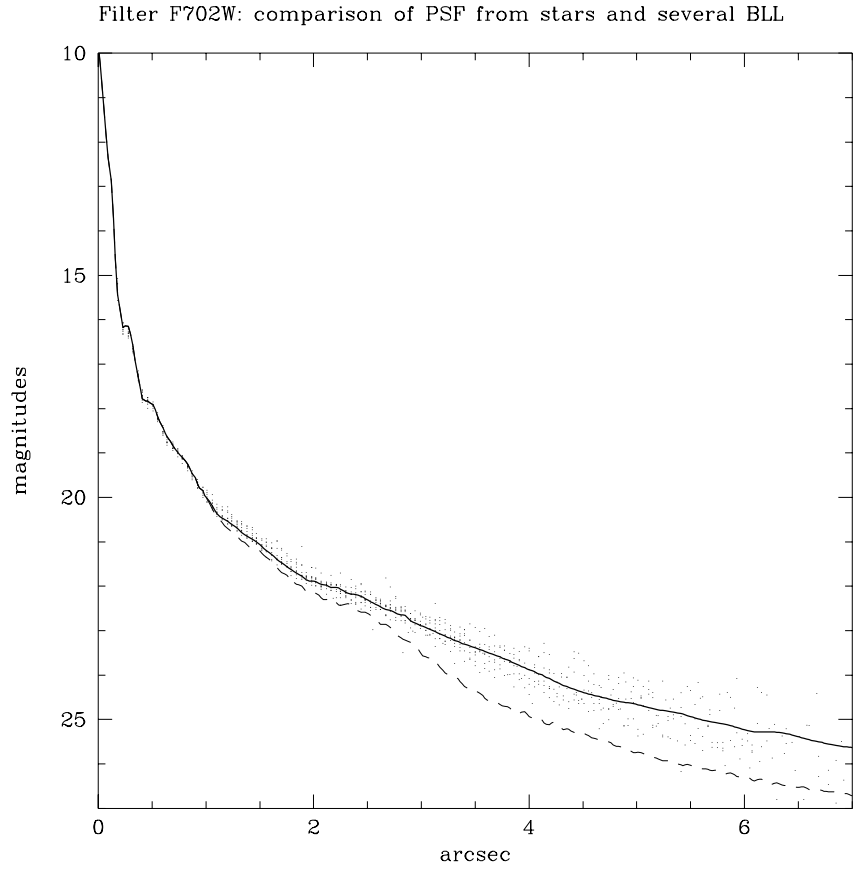,height=4.5in}
\caption{\footnotesize
Comparison of Tiny Tim PSF model (dashed line) with empirical profile
for several observed stars (dots). The analytic form used in the analysis
(solid line) is a sum of the Tiny Tim model within 2~arcsec
and the average stellar PSF
outside 2~arcsec. (See \S~2.4 for details.)
\normalsize}
\end{figure}

Since the PC images of stars are slightly under-sampled and often saturated 
(to make the faint wings of the PSF detectable),
we constructed a composite PSF using the Tiny Tim model within
the first 2~arcsec and the average of the observed stars at larger radii.
We find that the excess flux due to scattered light (above the Tiny Tim PSF) 
can be represented analytically for the F814W filter by:
\begin{equation}
S(r) = 5.85 \times 10^{-4} ~ I_{\rm PSF} ~
	e^{-0.42 r } ~,
\end{equation}
where $I_{\rm PSF}$ is the integrated flux of the Tiny Tim PSF 
and $r$ is the distance from the center in arcsec.
This extra term is clearly important for determining whether an object
is resolved or not, and also in cases where the PSF wings are an appreciable 
fraction of the light from the galaxy (e.g., 1308+326 and 1823+568).

\subsection{One-Dimensional Profile Fitting}
\label{ssec:onedim}

Properties of the BL~Lac host galaxies were derived both from azimuthally
averaged radial profiles and from two-dimensional surface photometry
(isophote fitting, \S~\ref{ssec:twodim}). 
The 1-D radial profiles were determined by
averaging the flux over annuli spaced at 1-pixel intervals. The 
statistical error
associated with the source flux in each annulus was computed from
the statistical noise, read noise, sky fluctuations,
and digitization noise.
We then added in quadrature an estimate of the systematic 
uncertainty in the PSF model, assumed to be 10\% of the 
point source contribution in that annulus.
We excluded the central 0.1~arcsec
to avoid the effects of undersampling, which can bias the fit strongly
at the very center. 
In some cases we combined annuli at large radii to increase
the signal-to-noise ratio.

The radial surface brightness profiles are shown in Figure~3. (Profiles of
the other three radio-selected BL~Lacs are given in Paper~I, and are
essentially unchanged by minor differences in the analysis procedure.)
We fitted the data with models consisting of PSF plus galaxy, all 
convolved with the PSF. That is, rather than subtracting the PSF first
then analyzing the galaxy light, we allowed the PSF normalization and the
galaxy brightness to vary simultaneously and independently, minimizing
chi-squared to find the best fit model. This avoids the {\it a priori}
determination of the PSF normalization, which can
introduce systematic bias in the derived galaxy luminosities.

\begin{figure}
\psfig{figure=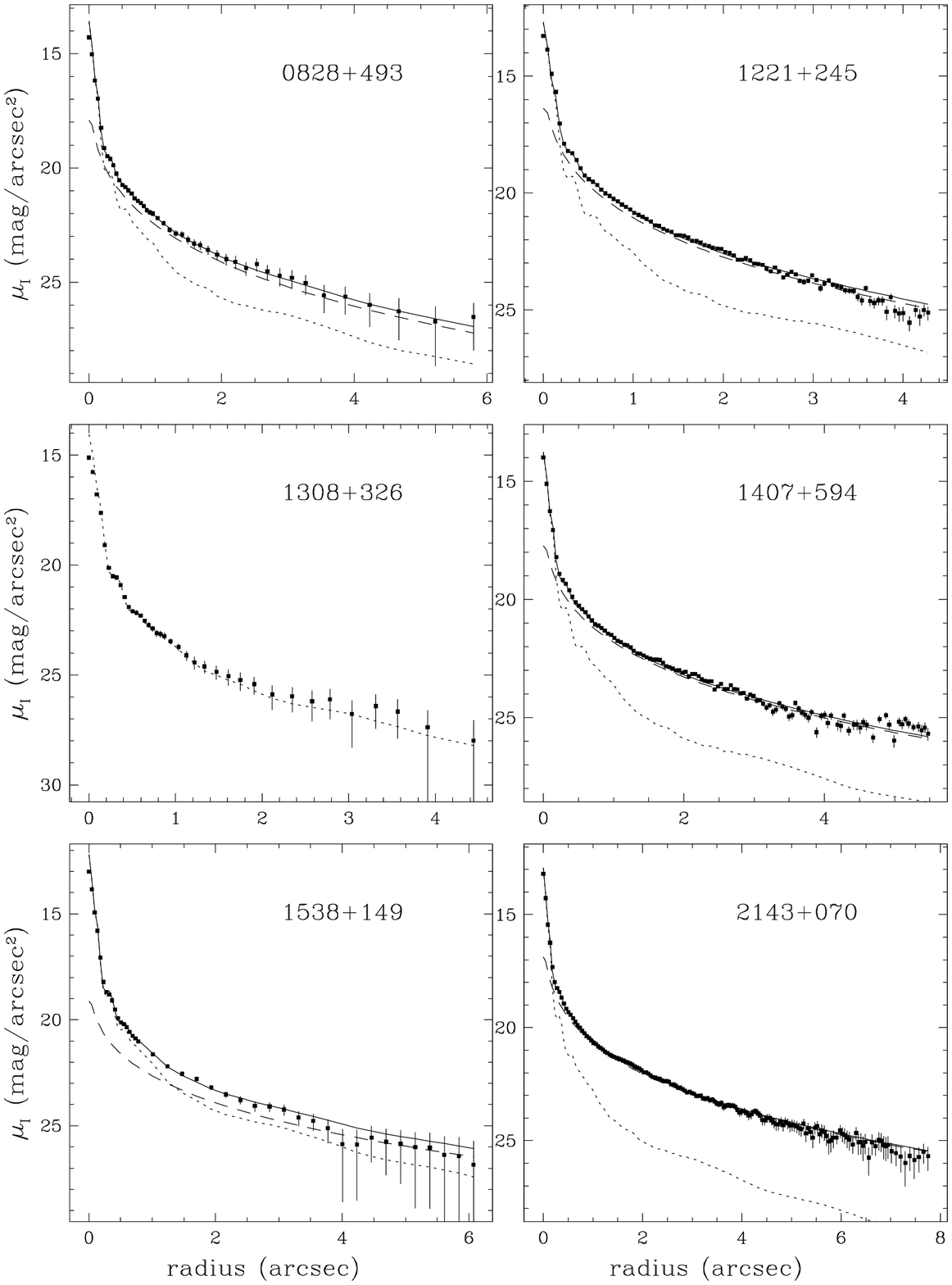,height=6.5in}
\caption{\footnotesize
Radial surface brightness profiles of the six new BL~Lac objects
(squares with error-bars), with best-fit model (solid line)
consisting of PSF (dotted line) plus galaxy (de~Vaucouleurs profile
convolved with PSF; dashed line). ``Red'' objects from our program
shown at left; our analysis of ``blue'' objects from Jannuzi
program at right. 
Profiles of the other three ``red'' objects in our program
appear in Paper~1 and are essentially unchanged.
\normalsize}
\end{figure}

We fitted each BL~Lac profile with the PSF alone and with the PSF plus
either an exponential disk or a de Vaucouleurs $r^{1/4}$ law. To decide
whether the host galaxy was resolved we compared, via an F-test,
the PSF-only and PSF-plus-galaxy fits; our threshold was that 
the galaxy is formally detected when the galaxy fit is better at
99\% confidence (we also require this to hold when the
sky is increased by $1\sigma$).
Similarly, we evaluated which of the two galaxy models was preferred
using an F-test at the same threshold. 

Given that we exclude the central 0.1~arcsec, we needed to bound the 
fitted point source to avoid unreasonably bright values. By evaluating 
several unresolved sources, we determined that the point source 
magnitude should be no more than 0.4~mag brighter than the total 
flux within the first 0.3~arcsec (where the flux is always dominated 
by the PSF), and so this limit was imposed on the fits.

The model fits were generally good, with no systematic deviation 
of points from the assumed model, but chi-squared values were
high. Based on similar random but excessive residuals
for $\sim 100$ {\it HST} snapshot observations of 
BL~Lac objects (Urry et al., in preparation), 
we conclude that chi-squared is high not because the model is
a poor representation of the data but because the errors are
systematically underestimated by $\sim \sqrt{2}$. 
Therefore we multiply the statistical errors by this correction
factor, following a procedure developed for {\it IUE} data
(discussed extensively by Urry et al. 1993).
In any case, this correction factor affects only the size of 
parameter errors determined from the chi-squared contours.

Results from the one-dimensional fitting are given in Table 2 and Figure~3
shows the best-fit model (solid line) for the PSF (dotted line) plus 
elliptical galaxy (dashed line).
In general, the best-fit PSF normalizations (i.e., nuclear magnitudes)
differ for the de~Vaucouleurs and disk model,
the derived value associated with the disk model being systematically 
brighter since it compensates for a relative lack of galaxy
central surface brightness.
Only when we can discriminate between the two galaxy models do we 
have an unambiguous estimate of the point source flux. 

We assessed statistical 
uncertainties in the fit parameters from multi-dimensional
chi-squared confidence contours, shown in Figure~4.
Quoted in Table~2 are the 68\% confidence uncertainties ($\Delta
\chi^2 = 2.3$ for the two parameters of interest, host galaxy magnitude
and half-light radius) from the box circumscribing the contour.
These represent statistical uncertainties only, while systematic errors
actually dominate the uncertainty (see 
\S~\ref{ssec:xcompare}).
Based on analysis of simulated data, 
the typical uncertainty in simply measuring the host galaxy magnitude,
independent of assumptions about PSF shape or normalization
(which increase the total systematic error), 
is $\Delta m \lesssim 0.2$~mag
when the point source brightness is within 2~mag of the host galaxy.

\begin{figure}
\psfig{figure=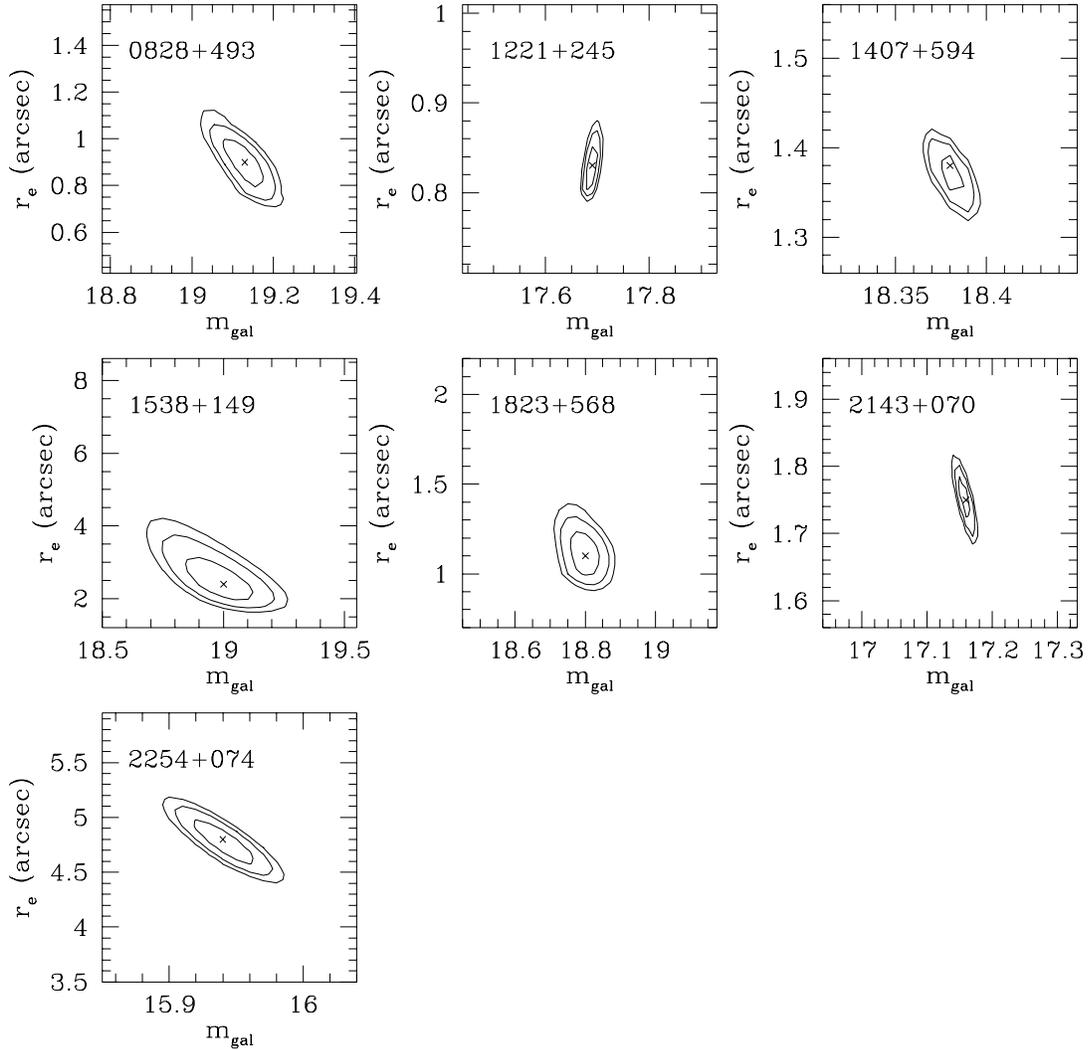,height=5.5in}
\caption{\footnotesize
Chi-squared confidence contours for the 
two parameters of interest (de~Vaucouleurs model fits), 
host galaxy magnitude ($M_{gal}$) and half-light radius
($r_e$). Uncertainties in Table~2 were derived from the 
68\% contours ($\Delta \chi^2 = 2.3$), the innermost contour
shown; also shown are the 95\% ($\Delta \chi^2 = 6$) and
99\% ($\Delta \chi^2 = 9.2$) contours.
The best fit (minimum $\chi^2$) is indicated with an x.
The limits of each box indicate the parameter space
explored in the fitting procedure.
\normalsize}
\end{figure}

For the two unresolved BL~Lac objects (0814+425, 1308+326), 
we determined 99\% confidence upper limits 
(statistical errors) to the host galaxy magnitudes 
($\Delta \chi^2 = 6.6$ for one parameter of interest, $M_{gal}$),
fixing the half-light radius at $r_e =10$~kpc, 
slightly larger (to be conservative)
than the weighted average for the seven resolved objects.

Published values of K corrections for elliptical galaxies, whether empirical
or theoretical, span a wide range of values. 
For example, recent determinations of B-band K corrections 
differ by up to 0.1~mag at $z=0.2$ and by up to $\sim0.5$~mag 
at $z\sim0.6$, the highest redshift relevant here 
(Kinney et al. 1996, Fukugita et al. 1995, Frei \& Gunn 1994, 
King \& Ellis 1985). Fortunately, the I-band K correction
is smaller and so are the uncertainties. Our values, listed in Table~2,
were obtained
from convolving observed elliptical galaxy spectra (Kinney et al. 1996),
redshifted appropriately, with the I-band filter response.
Comparing with values summarized in Fukugita et al. (1995) and with
Bruzual \& Charlot (1993) models convolved with the filter response,
we estimate the systematic uncertainties are 
$\lesssim 0.02$~mag at $z=0.2$, $\lesssim 0.1$~mag at $z=0.6$, and
$\sim 0.25$~mag at $z\sim1$.

\subsection{Color Profiles of Host Galaxies}
\label{ssec:colorprof}

For six of the nine BL~Lac objects, we have {\it HST} images 
in two filters (Table~1). The additional images,
either in the F702W (0828+493) or F606W filter (1407+595, 1538+149,
1823+568, 2143+070, 2254+074), are from our Cycle~6 snapshot survey 
(Urry et al., in preparation). Although they have shorter exposure
times and slightly worse signal-to-noise ratios than the pointed
observations (and only one roll angle), these data provide useful
color information for the host galaxies at the same high spatial 
resolution.

To compute the host color radial profile we first fitted the observed
radial profile with a PSF-plus-de Vaucouleurs model, then subtracted
the best-fit PSF model to obtain the galaxy profile in each filter. 
The difference is then the azimuthally averaged color profile, 
shown in Figure~5. 

\begin{figure}
\psfig{figure=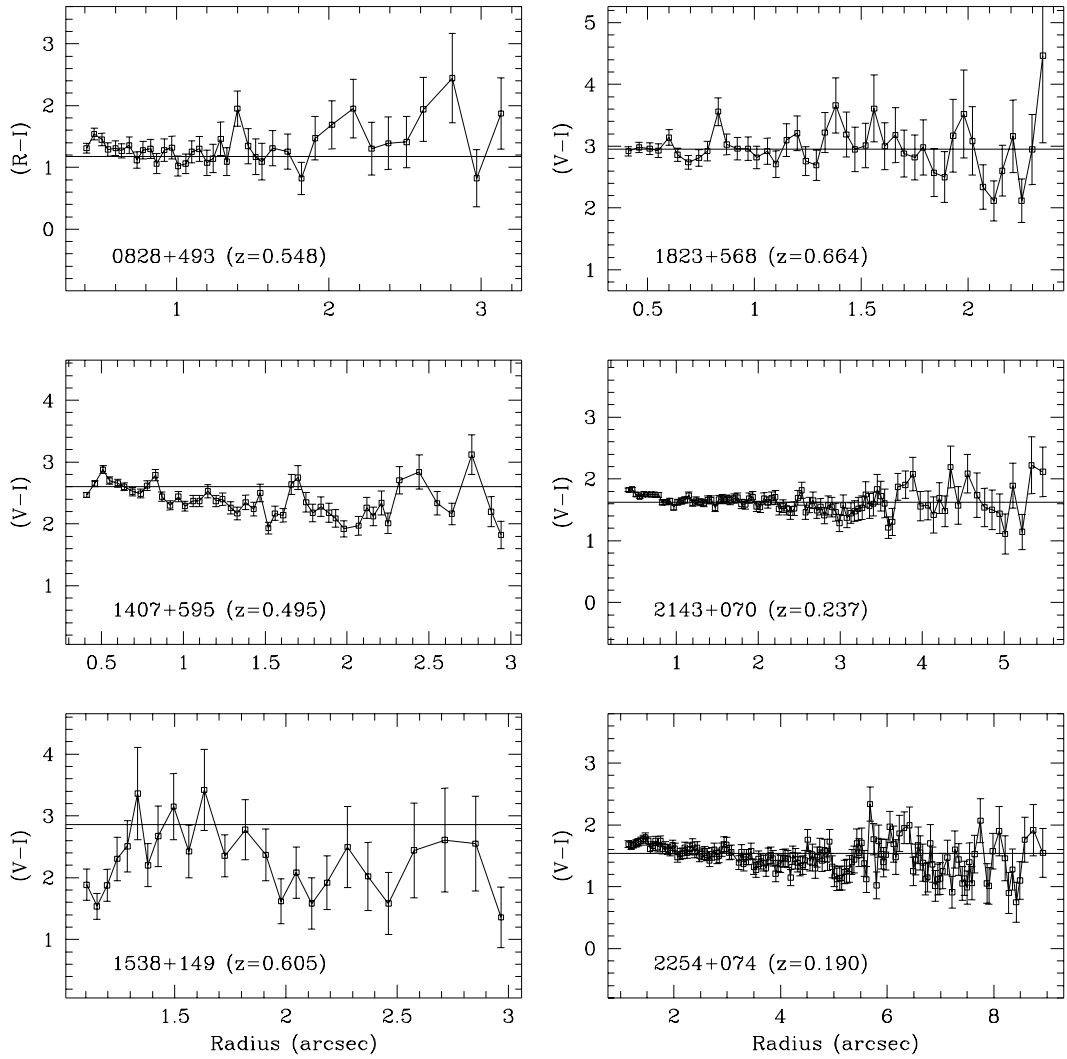,height=5.5in} 
\caption{\footnotesize
Color profiles for the six BL~Lacs with images in 
multiple WFPC2 filters. None shows a strong gradient outside
1~arcsecond. The horizontal line represents the expected color for a
normal elliptical galaxy at the redshift of the BL~Lac.
The profile of 1538+149 is bluer than expected, especially at the
center, but the bluer (R-band) image is the weakest detection of 
this set.
\normalsize}
\end{figure}

In general, there is little variation in color across the host galaxy at this
spatial resolution and S/N ratio. The galaxy colors are roughly as 
expected (horizontal line) for redshifted emission from a passively
evolving elliptical galaxy with an old stellar population
(rest-frame colors $R-I = 0.70$, $V-I= 1.31$).
The only marginal deviation from the expected color is
for 1538+149, which is slightly bluer than expected; however,
the host galaxy of this distant source ($z=0.605$) is only marginally
detected in the F606W snapshot image and therefore this 
result should be re-evaluated with better V-band data.

The galaxy colors are consistent with our finding that de Vaucouleurs 
models are generally better fits to the surface brightness profiles
than disks.
Determining whether there is any significant amount of recent star formation 
requires bluer images than we have. Based on the present data alone, 
the host galaxies appear dominated by old stars and
show no evidence of being strongly affected by the presence of a luminous
active nucleus.

\subsection{Two-Dimensional Surface Photometry}
\label{ssec:twodim}

We also did 2-dimensional surface photometry to probe the ellipticity,
isophotal twists, and point-source centering with respect to the host
galaxy. The 2-dimensional analysis was done as in Paper~I, using an
interactive numerical mapping package to derive isophotes 
(AIAP; Fasano 1994), which were fitted with ellipses down to $\mu_{I}$ =
25~mag~arcsec$^{-2}$ for each resolved object. There are four free
parameters per isophote --- intensity, center position, ellipticity,
and position angle --- which allow us to characterize the
morphological and the photometric properties of the galaxy. We
performed the surface photometry both on the full images and after
subtraction of a PSF scaled to match the flux in an annulus from 2 to
5 pixels, where the point source dominates but effects of
undersampling are reduced. The results are consistent for the outer
isophotes, where the galaxy is dominant.

The derived ellipticities are in general very small, $\epsilon \lesssim
0.1$ (Table~3), reflecting the obvious roundness of the images
(Fig.~1) and justifying our 1-dimensional approach. Averaging over
annuli maximizes the signal-to-noise ratio, allowing us to go deeper
in the outer regions of the galaxy, and is computationally simpler
than 2-D surface photometry, which is important for determining errors
because of the many convolutions involved. The radial surface
brightness profiles obtained from the 2-D analysis are in excellent
agreement with those derived from the azimuthally averaged profiles
taking into account the ellipticities.

We have investigated possible offsets of the
galaxy isophotes, which can be indicative of recent mergers or tidal
interactions. In Table~3 we give the normalized displacement $\delta
\equiv \sqrt{(X_c-X_o)^2 +(Y_c-Y_o)^2}/R$, where $X_c, Y_c$ are the
centers of isophotal ellipses, $X_o, Y_o$ is the center of the
innermost PSF-subtracted isophote, 
and $R$ is the radial distance to the particular elliptical
isophote; $\delta$ represents the variation of the position of the
center of each isophote with respect to the innermost. A systematic
trend of increasing $\delta$ was found for 1407+594, indicating that this
galaxy may have suffered some sort of interaction in the recent past. In
any case, the small displacement of isophotal centers in these host
galaxies is consistent with that found in normal ellipticals (e.g.,
Sparks et al. 1991; see also Colina \& De Juan 1995).

The comparison of the central point source position with respect to 
the host galaxy is relevant for the microlensing hypothesis
(Ostriker \& Vietri 1985). We therefore compared the 
point source location to the galaxy center,
defined by averaging the centers of all isophotes
unaffected by the central point source but bright
enough to be unaffected by strong statistical errors.
In all cases we can place a strong upper limit
to any off-centering of less than 1 pixel ($< 0.05$~arcsec), 
which argues against microlensing playing an important role in 
this BL~Lac sample.

\begin{deluxetable}{ccccc}
\small
\tablewidth{4.0in}
\tablenum{3}
\tablecaption{Surface Photometry of BL Lac Host}
\tablehead{
\colhead{Object} & \colhead{Ellipticity} & \colhead{PA\tablenotemark{(a)}} &
\colhead{$\langle \delta \rangle$} & \colhead{De-Center} \\
 \colhead{} & \colhead{} & \colhead{($\circ$)} & \colhead{} & \colhead{(arcsec)}
}
\startdata
0828+493  & 0.1     & \en170          & 0.03 & $<0.01$           \nl
1221+245  & 0.1     & \en165          & 0.03 & $<0.01$           \nl
1407+594  & 0.1-0.2 & \en\en\en30\en  &~~~~0.08$^{\rm (b)}$ & $<0.03$\nl
1538+149  & 0.1     & \en\en120\en    & 0.05 & $<0.03$           \nl
1823+568  & 0.1     & $\sim45$        & 0.05 & $<0.03$           \nl
2143+070  & 0.2-0.3 & \en\en\en56\en  & 0.01 & $<0.01$           \nl
2254+074  &$<$0.02\tablenotemark{(c)} & \nodata & 0.02 & $<0.05$ \nl
\enddata
\tablenotetext{(a)}{Position angle from north toward east.}
\tablenotetext{(b)}{Continuous (increasing) trend of $\delta$.}
\tablenotetext{(c)}{Ellipticity = 0.1 at $0.5< r<1$ arcsec.}
\end{deluxetable}
\normalsize

\section{HST Results and Comparison with Previous Analyses and
Ground-Based Imaging}
\label{sec:results}

\subsection{Properties of the BL~Lac Host Galaxies }
\label{ssec:properties}

Host galaxies are detected in seven of the nine BL~Lac objects. They
are generally very luminous and have surface brightness profiles
consistent with a de Vaucouleurs $r^{1/4}$ model, which is preferred
over a disk model in 6 of the 7 cases (and the seventh is
indeterminate). The average absolute magnitude is $\langle M_I
\rangle = -24.6 \pm 0.7$~mag (rms dispersion). Values of point
source magnitude, host galaxy magnitude, and half-light radius
are given in Table~2. For comparison, the brightest ellipticals 
in the nearby well-studied clusters 
Virgo (Caon, Capaccioli \& D'Onofrio 1994) 
and Coma (Jorgensen \& Franx 1994) have similar 
or slightly brighter absolute magnitudes, 
with $M_I \sim -24.9$~mag (for $H_0 = 50$~km~s$^{-1}$~Mpc$^{-1}$
and typical elliptical color $B-I = 2.2$).

The objects with detected host galaxies, in addition to 1823+568 and
2254+074 of Paper~I, are the ``red'' BL Lacs 0828+493
($z=0.548$) and 1538+149 ($z=0.605$), and all three ``blue'' BL Lacs
1221+245 ($z=0.218$), 1407+595 ($z=0.495$), and 2143+070 ($z=0.237$).
Two radio-selected objects, 0814+425 ($z\sim 0.25$ or 
$\sim1.25$?, Paper~I) and
1308+326 ($z=0.997$) remain unresolved even at {\it HST} resolution.
The images in the lower panel in Figure~1 show the three new
radio-selected BL Lac objects with the best-fit point source subtracted.

Our results (Tables~2 and 3) are largely in agreement with previous
surveys of BL~Lac objects, which found that with a few exceptions, the
host galaxies are luminous, round ellipticals (Abraham, McHardy \&
Crawford 1991; Stickel et al. 1993; Wurtz, Stocke \& Yee 1996, hereafter WSY; 
Falomo 1996). Because we probe higher redshifts on average
than these ground-based surveys, it is not surprising that the
detected host galaxies are also more luminous on average, with I-band
luminosities and sizes comparable to brightest cluster galaxies
(Taylor et al. 1996; WSY) and to Fanaroff-Riley type~I radio galaxies
(Ledlow \& Owen 1996; Urry et al., in preparation; cf. WSY). Like
WSY, we see no significant difference in the host galaxies of ``red''
or ``blue'' BL~Lacs, although the present sample is quite small.

The colors of the host galaxies are typical of the old stellar
populations in ellipticals (Fukugita et al 1995; Bressan et al. 1994).
Figure~6 shows the observed $V-I$ colors of the five host galaxies
with {\it HST} images in both the F606W and F814W filters. A sixth
galaxy observed in F702W and F814W is also plotted (triangle) assuming
$V-R=1.4$~mag (valid for an elliptical at z=0.5; Fukugita et al. 1995).
Redder and/or bluer images would greatly improve our constraints on
the stellar populations; in particular, we have no constraints on very
recent star formation from these red images. 

\begin{figure}
\psfig{figure=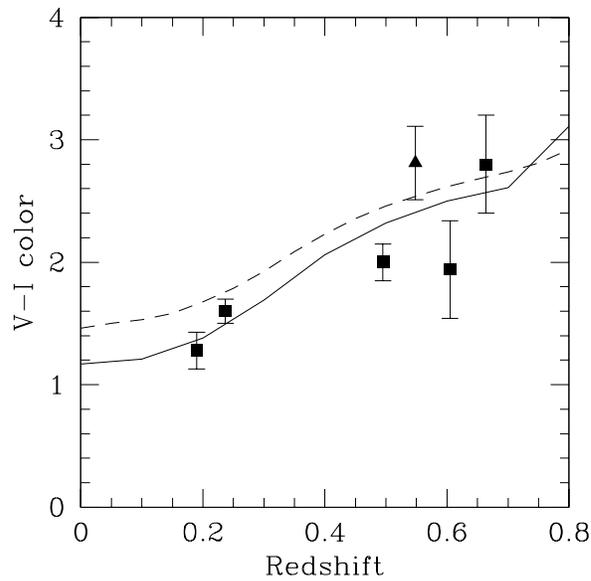,height=3.0in} 
\caption{\footnotesize
Average observed colors of BL~Lac host galaxies compared 
with those expected from old stellar populations as a function of
redshift. {\it Squares:} observed V--I colors;
{\it triangle:} converted from R--I (see text).
The solid and dashed lines indicate the
expected color for normal elliptical galaxies assuming passive
evolution and $q_0=0$ as reported by Fukugita et al. (1995) and
Bressan et al. (1994), respectively.
\normalsize}
\end{figure}

Below we discuss individual objects in more detail.

\subsection{0828+493 (z=0.548)}
\label{ssec:0828}

The host galaxy is clearly detected in the {\it HST} image and is well
centered on the point source. The nebulosity shows slightly elongated
isophotes, with $\epsilon\sim 0.1$ and P.A.$ \sim 170^\circ$ (E of N).
The luminosity profile is significantly better fitted by a de
Vaucouleurs law than by an exponential disk. The best-fit host galaxy
has a K corrected absolute I-band magnitude of $M_I=-24.6$~mag and a
half-light radius of $r_e = 7.6$~kpc. The $R-I$ color profile of
0828+493 (Fig. 5) is basically flat and consistent with the expected
color of an elliptical galaxy at this redshift, $R-I = 1.2$~mag.

The host galaxy was marginally resolved from the ground by WSY, who
report for a de~Vaucouleurs law fit a galaxy magnitude (Gunn system)
of $m_r = 20.57$~mag, which corresponds to $m_I=19.1$~mag,\footnote{We
assume an intrinsic color $r-R=0.30$, a value that is correct to
within 0.1~mag at all redshifts considered here (Fukugita et
al. 1995).}
fully consistent with our determination, as is their
half-light radius, $r_e = 0.7$~arcsec
(compared to $0.9 \pm 0.2$~arcsec). The WSY data were insufficient to
distinguish between an exponential disk or de Vaucouleurs profile.

\subsection{1308+326 (z=0.997)}
\label{ssec:1308}

The host galaxy of this high redshift object is not resolved in our
WFPC2 image even though Stickel et al. (1993) reported detection of
extended isophotes in a ground-based R-band image, specifically from
radii $\sim5$ to $\sim10$~arcsec, at surface brightnesses from $\mu_R
\sim 27$-29~mag~arcsec$^{-2}$ (Stickel 1990). The total integration
time for the {\it HST} I-band image is three times that of the Stickel
et al. R-band image, and we also have a deeper R-band image taken with
the NOT in 0.8-arcsec seeing (Falomo et al. 1996). We do detect
clearly the galaxy Stickel et al. saw $\sim5$~arcsec to the southwest
but no significant extension in the outer isophotes. Any nebulosity
surrounding the BL Lac nucleus must be fainter than $\mu_I >
26$~mag~arcsec$^{-2}$, and the host galaxy fainter than $m_I{\rm
(galaxy)} > 20.1$~mag, or $M_I > -26.4$~mag. Due to its high
redshift, this object has been suggested as a case of microlensing by
stars in a foreground galaxy.

\subsection{1538+149 (z=0.605)}
\label{ssec:1538}

The host galaxy of this object is resolved with {\it HST} but the
surface brightness is low and it is the only one of the seven for
which we can not uniquely determine its morphology. The luminosity
profile is equally well fitted with an exponential disk or a
de~Vaucouleurs law. For a de~Vaucouleurs model, the galaxy has 
$M_I=-25.1$~mag and $r_e = 21$~kpc (with large uncertainties), 
indicating the host
is both large and luminous. The $V-I$ color profile (Fig. 5) is
essentially flat and roughly consistent with the expected color, $V-I
= 2.85$, of an aging elliptical galaxy at this redshift. The galaxy
is relatively round, with ellipticity $\epsilon\sim 0.1$, and well
centered on the point source.

There are several faint galaxies within $\sim$50~kpc of the BL~Lac.
Based on their sizes and magnitudes, these are consistent with $L^*$
galaxies at the redshift of the BL~Lac object. Unfortunately, we are
unable to determine their colors since they are not visible in the
shorter F702W snapshot observation.

This BL~Lac object was marginally resolved by WSY, who reported a
galaxy magnitude of $m_r = 19.9$~mag, corresponding to $m_I=18.3$~mag,
2.5 sigma brighter than our best fit. Their half-light radius, $r_e =
1.5$~arcsec, is consistent with ours within the uncertainties.

\subsection{Reanalysis of HST Data for 0814+425, 1823+568, and 2254+074}
\label{ssec:oldrbl}

We have reanalyzed the three sources reported in Paper I 
(0814+425, 1823+568, 2254+074), using the modified PSF and 
flux calibration. This yields consistent results for 0814+425
and 2254+074 when the slight offset (0.1~mag) in the zero point of the 
calibration (Eq. 2) is taken into account. 
For 1823+568, the revised host galaxy magnitude 
is 0.5 mag fainter than previously reported, primarily due
to the new PSF adopted (also to slightly modified sky background,
calibration, and fitting procedure).
This galaxy was detected well above the PSF, suggesting that the
effect on the derived magnitude could be much larger for fainter
host galaxies. Comparisons to other measurements of these three
BL~Lac objects were given in Paper~I.

\subsection{Reanalysis of HST Data for X-Ray Selected BL~Lacs 
\label{ssec:xbl}
(1221+245, 1407+595, 2143+070)}
\label{ssec:xcompare}

We retrieved from the {\it HST} archive the WFPC2 images of 1221+245,
1407+595, and 2143+070, three ``blue'' BL~Lac
objects from the Cycle~5 program of J97. Their observations and ours
were designed to be similar in filter and depth so that the two data
sets could be compared easily. J97 used a different flux calibration 
and followed a different analysis
procedure, first subtracting the PSF with a normalization value based
on visual inspection of the residuals, and then fitting the residual
light with a galaxy model. In our analysis, the two components (PSF
and galaxy) are fitted simultaneously.

The results reported here (Table~2) differ slightly from those given
by J97. In all cases we confirm that the host is clearly resolved and
that a disk model for the host is ruled out but the values for the
galaxy magnitudes are in two cases slightly discrepant. 
These differences stem from slightly different analysis procedures;
because neither approach can be deemed better than the other,
this allows us to estimate the systematic uncertainty 
in the reported values.

For 2143+070 the agreement is excellent: our values are $m_I=17.16 \pm
0.02$~mag and $r_e = 1.8 \pm 0.1$~arcsec, compared to $m_I=17.13$ and
$r_e = 1.79$~arcsec from J97. However, for 1407+595 we found
$m_I=18.38 \pm 0.02$~mag and $r_e = 1.38 \pm 0.04$~arcsec, compared to
$m_I=18.53$~mag and $r_e = 1.5$~arcsec; the difference of $\Delta m_I
= 0.15$~mag is well outside our estimated statistical uncertainty.

For 1221+245 the difference is even larger: we obtained $m_I=17.69 \pm
0.03$~mag and $r_e=0.83 \pm 0.04$~arcsec while J97 found $m_I=17.41$
and $r_e = 0.6$~arcsec, a difference of $\Delta m= 0.28$~mag. In this
case we note that the best-fit galaxy reported by J97 appears to be
systematically brighter than the observed data points after
subtraction of the PSF. This, together with the presence of a residual
PSF modulation in the PSF-subtracted profile (see their Fig. 5),
suggest that their galaxy luminosity is slightly overestimated. This
object was also marginally resolved by WSY from the ground; using a procedure
similar to ours they found $m_r = 18.65$~mag, corresponding to $m_I =
17.66$~mag, a value consistent with ours.

Larger discrepancies are found between the derived magnitudes of
the point sources, with our values being systematically brighter (by
0.4 to 0.9 mag) than those reported by J97. This is due in part to
our excluding the central 0.1~arcsec, so that our constraints are
weaker and the associated uncertainties larger (Table~2). 
(We deliberately use loose constraints on the fitted PSF because 
the model is not perfect.)
Also, the point sources in these objects are weak compared to the galaxy,
much more so than for the ``red'' BL Lac objects, so one would expect
relatively larger uncertainties in their measured magnitudes.

\subsection{Systematic Errors}
\label{ssec:systematic}

The above comparison indicates that measurements of the host galaxy
magnitudes can be uncertain by 0.2-0.3~mag, even for bright, easily
detected elliptical galaxies like 1221+245. One obvious source of
systematic error is uncertainty in the PSF, which we incorporated in
the statistical error in an {\it ad hoc} way (\S~\ref{ssec:onedim}).
It is this uncertainty, combined with variations in how the PSF
normalization is estimated and how PSF wings are described, that
leads to the systematic discrepancies of up to $\sim 0.5$~mag in the
fitted host galaxy magnitudes described in the previous two sections.

Another source of systematic error is the uncertainty in the sky
background, which affects the surface brightness profile most strongly
in the outer regions of the host galaxy. This is usually not critical,
as the profile is not weighted heavily there and the signal is
averaged over a large area, repressing all but the largest scale
fluctuations. When the galaxy is only marginally resolved, however,
the sky fluctuation can be a much more significant source of error.
Overestimating the sky by 0.5$\sigma$ can change the derived host
galaxy magnitude by up to +0.2~mag and reduce the half-light radius by
0.4~arcsec; underestimating by $0.5\sigma$ has a larger effect,
changing $M_{gal}$ by up to -0.4~mag and increasing the half-light
radius by up to 2~arcsec. Under-subtracting the background, however,
leaves a characteristic signature whereby the outer radial surface
brightness profile is parallel to the sky level, so we are
unlikely to have made an error of this magnitude. 

In addition,
conversion to absolute magnitude introduces a systematic error because
the K correction is not precisely known. The K corrections in the literature
vary widely; for elliptical galaxies, published B-band values differ
by 0.1~mag at $z\lesssim0.2$ and as much as 0.5~mag at $z\sim1$
(\S~\ref{ssec:onedim}). The range of values for spiral galaxies is
even larger. The difference between the K corrections for
E-type and Sb-type spectra is of course still larger (as much as 2~mag
at $z\sim1$). Even if the adopted galaxy type is correct, published
values of absolute magnitudes of host galaxies can differ because
different K corrections were applied. We used the best available data 
for elliptical galaxy spectra (Kinney et al. 1996)
and estimated uncertainties by comparing to a widely used recent model
(Bruzual \& Charlot 1993).
The K corrections we used are given in Table~2.

Finally, even starting from the same {\it HST} data, derived galaxy
properties in the literature can differ mainly because of different
fitting assumptions, but also because of different calibrations,
different aperture sizes for photometry, as well as the usual
differences in cosmology. It is not trivial to compare final absolute
magnitudes of host galaxies from different authors.

\section{Discussion and Conclusions}
\label{sec:disc}

We have shown that with {\it HST} it is easy to detect and
characterize the host galaxies of low-luminosity AGN like BL~Lac
objects up to moderately high redshifts, $z \sim 0.7$. We detected
host galaxies in seven of the nine BL~Lac objects in our combined
sample, including the highest redshift and most luminous host galaxy
ever detected in a BL~Lac object (1823+568 at $z=0.664$). Our
measurements of the host galaxies of three more BL~Lac objects at
redshifts $z\gtrsim 0.5$ give improved estimates of their luminosities
and surface brightness profiles compared to previous ground-based
measurements by WSY (one of which, 1538+149, was a marginal detection).

The detected host galaxies are luminous ellipticals. Their
average K corrected absolute magnitude is $\langle M_I \rangle =
-24.6$~mag, more than 1~mag brighter than $L^*_I =-23.3$~mag
(Mobasher, Sharples \& Ellis 1993). This is the same as for brightest
cluster galaxies ($M_I = -24.6$~mag; Thuan \& Puschell 1989) and for
Fanaroff-Riley type~I radio galaxies (Ledlow \& Owen 1996), which are
often found in moderate to rich cluster environments. Indeed, the
extended (unbeamed) radio luminosities of BL~Lac objects (Urry \&
Padovani 1995), combined with their luminous host galaxies, strongly
support the unification picture with FR~I galaxies constituting the
parent population. Further refinement of this picture is possible from
consideration of the larger environments of the BL~Lacs (Stickel et
al. 1993; Falomo, Pesce, \& Treves 1993, 1995; Pesce, Falomo, \&
Treves 1994, 1995; Smith, O'Dea \& Baum 1995; Wurtz et al. 1997).

As well as being luminous, the host galaxies of BL~Lac objects are
large. We have too few objects for a quantitative comparison with
samples of radio galaxies, quasars, or cluster ellipticals, but
qualitatively the half-light radii are comparable to those of
brightest cluster galaxies and are consistent with the larger, lower
surface brightness end of the $\mu_e$-$r_e$ projection of the
fundamental plane (Djorgovski \& Davis 1987, Hamabe \& Kormendy 1987).

In 6 of the 7 BL~Lac host galaxies, the de Vaucouleurs $r^{1/4}$ model
gives a significantly better fit to the surface brightness profile
than an exponential disk model. (In the seventh, both fits are good
and neither is preferred.) Similarly, the colors are as expected for
passively evolving elliptical galaxies with old stellar populations at
the appropriate redshift. To see signs of recent star formation,
however, requires bluer images than we currently have.

Although our sample is small, we see no differences between the host
galaxies of ``red'' and ``blue'' BL~Lac objects. The ``red'' objects,
which have systematically higher bolometric luminosities, may be
redder because the highest energy electrons cool quickly by scattering
ambient UV and X-ray photons to gamma-ray energies (Ghisellini et
al. 1998). The ``blue'' objects have less luminous nuclei and jets
with lower kinetic powers (Celotti, Padovani \& Ghisellini 1997).
These two classes of BL~Lac object therefore reflect two different
kinds of jets (probably extrema of a continuous distribution), which
in turn indicates different jet formation and/or evolution. Our
results on the host galaxies (see also WSY) suggest that nuclear
properties, which can strongly influence jet formation and
propagation, do not have a dramatic effect on the host galaxy (or
vice-versa).

With the high spatial resolution of {\it HST}, we are able to place
tight limits on any de-centering of the BL~Lac nucleus. If any of our
BL~Lac objects were actually background quasars microlensed by stars
in a foreground galaxy (which we are calling the host galaxy), there
could well be an offset between the position of the nucleus (the
amplified background quasar) and the lensing galaxy. Instead, the
nuclei are well-centered in the host galaxy, with deviations typically
less than 0.03~arcsec. For the ``blue'' BL~Lac object 1407+595, the
nucleus is also well centered with respect to the inner isophotes, but
the galaxy isophotes shift in a way seen in many elliptical galaxies,
possibly indicating an interaction or merger (Sparks et al. 1991,
Colina \& De Juan 1995).

The present sample of nine well-studied BL~Lac objects indicates 
the clear trends described above:
\begin{enumerate}
\item
The host galaxies of BL Lac objects out to $z\sim0.7$
are well resolved and characterized with {\it HST}.
\item
Elliptical galaxy profiles are strongly preferred to disk profiles,
and the colors are consistent with aging stellar populations.
\item
The galaxies are large, round, and roughly one magnitude
more luminous than $L^*$.
\item
There is no significant difference between the host galaxies of 
``red'' and ``blue'' BL~Lac objects, despite their significantly
different radio jets.
\item
We do not detect the de-centering signature expected if BL~Lac objects
are background quasars microlensed by stars in the ``host'' (actually,
foreground) galaxy.
\end{enumerate}

We can establish these trends for the BL~Lac class with the larger
samples now available. More than 100 BL~Lac objects have been
observed with {\it HST} in our Cycle~6 snapshot program, albeit to
less depth, and host galaxies have been detected in more than half,
including nearly 90\% of those with redshifts $z\lesssim0.5$ (Urry et
al., in preparation). Comparing to radio galaxies over a similar
redshift range, we can investigate unified schemes. Should they be
confirmed, properties of the BL~Lac host galaxies and near
environments are basically universal to all low-luminosity radio-loud
AGN. We can then explore the larger sample of all radio-loud AGN for
trends in luminosity and redshift.

\acknowledgements
Support for this work was provided by NASA through grant numbers
GO5938.01-94A and GO5939.01-94A from the Space Telescope Science
Institute, which is operated by AURA, Inc., under NASA contract
NAS~5-26555. MG acknowledges support from the Hubble Fellowship
Program through grant number HF 1071.01-94A, awarded by the Space
Telescope Science Institute. We thank Chris Burrows, Marcella Carollo,
Bob Hanisch, Eric Hooper, John Hutchings, Buell Jannuzi, John Krist, and Brad
Whitmore for useful and stimulating discussions. RF thanks the {\it
HST} visitor program for hospitality during several visits to
STScI. We thank G. Fasano for the use of the AIAP package. This
research made use of the NASA/IPAC Extragalactic Database (NED),
operated by the Jet Propulsion Laboratory, Caltech, under contract
with NASA, and of NASA's Astrophysics Data System Abstract Service (ADS).

\newpage
\begin{figure}
\psfig{file=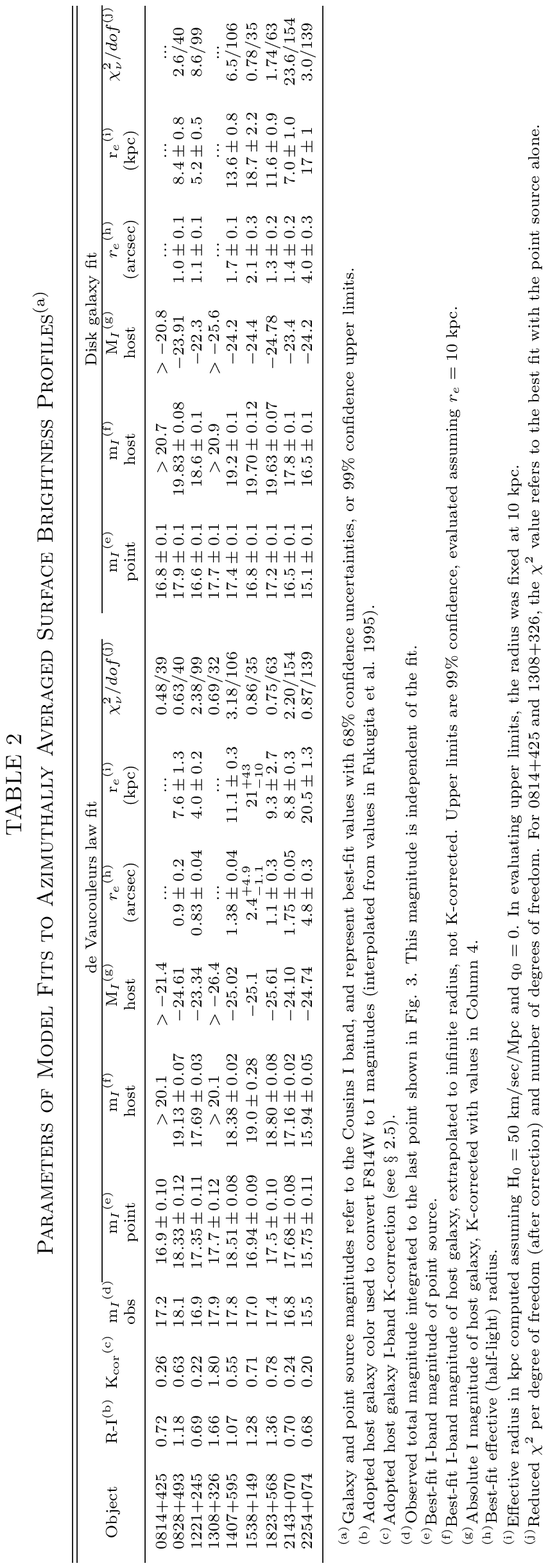,angle=180}
\end{figure}

\end{document}